\input harvmac

\input amssym

\def\unit{\relax{\rm 1\kern-.26em I}}
\def\nada{\relax{\rm 0\kern-.30em l}}
\def\tilde{\widetilde}


\def\CX{{\cal X}}

\def\det{{\rm det}}

\noblackbox
\def\IL{\relax{\rm I\kern-.18em L}}
\def\IH{\relax{\rm I\kern-.18em H}}
\def\IR{\relax{\rm I\kern-.18em R}}
\def\IC{\relax\hbox{$\inbar\kern-.3em{\rm C}$}}
\def\IZ{\relax\ifmmode\mathchoice
{\hbox{\cmss Z\kern-.4em Z}}{\hbox{\cmss Z\kern-.4em Z}} {\lower.9pt\hbox{\cmsss Z\kern-.4em Z}}
{\lower1.2pt\hbox{\cmsss Z\kern-.4em Z}}\else{\cmss Z\kern-.4em Z}\fi}

\def\CH {{\cal H}}

\def\CS {{\cal S}}


\def\CW{{\cal W}}

\def\CS {{\cal S }}

\def\det{{\rm det}}
\def\Tr{{\rm Tr}}

\font\manual=manfnt \def\dbend{\lower3.5pt\hbox{\manual\char127}}

\def\IZ{\relax\ifmmode\mathchoice
{\hbox{\cmss Z\kern-.4em Z}}{\hbox{\cmss Z\kern-.4em Z}} {\lower.9pt\hbox{\cmsss Z\kern-.4em Z}}
{\lower1.2pt\hbox{\cmsss Z\kern-.4em Z}}\else{\cmss Z\kern-.4em Z}\fi}

\def\bar{\overline}
\def\CS{{\cal S}}
\def\CH{{\cal H}}

\def\rt2{\sqrt{2}}
\def\irt2{{1\over\sqrt{2}}}

\def\slashchar#1{\setbox0=\hbox{$#1$}           
   \dimen0=\wd0                                 
   \setbox1=\hbox{/} \dimen1=\wd1               
   \ifdim\dimen0>\dimen1                        
      \rlap{\hbox to \dimen0{\hfil/\hfil}}      
      #1                                        
   \else                                        
      \rlap{\hbox to \dimen1{\hfil$#1$\hfil}}   
      /                                         
   \fi}

\def\foursqr#1#2{{\vcenter{\vbox{
    \hrule height.#2pt
    \hbox{\vrule width.#2pt height#1pt \kern#1pt
    \vrule width.#2pt}
    \hrule height.#2pt
    \hrule height.#2pt
    \hbox{\vrule width.#2pt height#1pt \kern#1pt
    \vrule width.#2pt}
    \hrule height.#2pt
        \hrule height.#2pt
    \hbox{\vrule width.#2pt height#1pt \kern#1pt
    \vrule width.#2pt}
    \hrule height.#2pt
        \hrule height.#2pt
    \hbox{\vrule width.#2pt height#1pt \kern#1pt
    \vrule width.#2pt}
    \hrule height.#2pt}}}}
\def\psqr#1#2{{\vcenter{\vbox{\hrule height.#2pt
    \hbox{\vrule width.#2pt height#1pt \kern#1pt
    \vrule width.#2pt}
    \hrule height.#2pt \hrule height.#2pt
    \hbox{\vrule width.#2pt height#1pt \kern#1pt
    \vrule width.#2pt}
    \hrule height.#2pt}}}}
\def\sqr#1#2{{\vcenter{\vbox{\hrule height.#2pt
    \hbox{\vrule width.#2pt height#1pt \kern#1pt
    \vrule width.#2pt}
    \hrule height.#2pt}}}}

\def\figin{\epsfcheck\figin}\def\figins{\epsfcheck\figins}
\def\epsfcheck{\ifx\epsfbox\UnDeFiNeD
\message{(NO epsf.tex, FIGURES WILL BE IGNORED)}
\gdef\figin##1{\vskip2in}\gdef\figins##1{\hskip.5in}
\else\message{(FIGURES WILL BE INCLUDED)}%
\gdef\figin##1{##1}\gdef\figins##1{##1}\fi}
\def\DefWarn#1{}
\def\figinsert{\goodbreak\midinsert}
\def\ifig#1#2#3{\DefWarn#1\xdef#1{fig.~\the\figno}
\writedef{#1\leftbracket fig.\noexpand~\the\figno}%
\figinsert\figin{\centerline{#3}}\medskip\centerline{\vbox{\baselineskip12pt \advance\hsize by
-1truein\noindent\footnotefont{\bf Fig.~\the\figno:\ } \it#2}}
\bigskip\endinsert\global\advance\figno by1}


\lref\CsakiSR{
  C.~Csaki, A.~Falkowski, Y.~Nomura and T.~Volansky,
  ``A New Approach to mu-Bmu,''
  arXiv:0809.4492 [hep-ph].
}
\lref\DvaliCU{
  G.~R.~Dvali, G.~F.~Giudice and A.~Pomarol,
  ``The $\mu$-Problem in Theories with Gauge-Mediated Supersymmetry Breaking,''
  Nucl.\ Phys.\  B {\bf 478}, 31 (1996)
  [arXiv:hep-ph/9603238].
}

\lref\IntriligatorCP{
  K.~A.~Intriligator and N.~Seiberg,
  ``Lectures on Supersymmetry Breaking,''
  Class.\ Quant.\ Grav.\  {\bf 24}, S741 (2007)
  [arXiv:hep-ph/0702069].
}

\lref\DineXK{
  M.~Dine, Y.~Nir and Y.~Shirman,
  ``Variations on minimal gauge mediated supersymmetry breaking,''
  Phys.\ Rev.\  D {\bf 55}, 1501 (1997)
  [arXiv:hep-ph/9607397].
}

\lref\DvaliCU{
  G.~R.~Dvali, G.~F.~Giudice and A.~Pomarol,
  ``The $\mu$-Problem in Theories with Gauge-Mediated Supersymmetry Breaking,''
  Nucl.\ Phys.\  B {\bf 478}, 31 (1996)
  [arXiv:hep-ph/9603238].
}

\lref\GiudiceCA{
  G.~F.~Giudice, H.~D.~Kim and R.~Rattazzi,
  ``Natural mu and Bmu in gauge mediation,''
  Phys.\ Lett.\  B {\bf 660}, 545 (2008)
  [arXiv:0711.4448 [hep-ph]].
}

\lref\MurayamaGE{
  H.~Murayama, Y.~Nomura and D.~Poland,
  ``More Visible Effects of the Hidden Sector,''
  Phys.\ Rev.\  D {\bf 77}, 015005 (2008)
  [arXiv:0709.0775 [hep-ph]].
}

\lref\GiudiceCA{
  G.~F.~Giudice, H.~D.~Kim and R.~Rattazzi,
  ``Natural mu and Bmu in gauge mediation,''
  Phys.\ Lett.\  B {\bf 660}, 545 (2008)
  [arXiv:0711.4448 [hep-ph]].
}

\lref\gmreview{
  G.~F.~Giudice and R.~Rattazzi,
  ``Theories with gauge-mediated supersymmetry breaking,''
  Phys.\ Rept.\  {\bf 322}, 419 (1999)
  [arXiv:hep-ph/9801271].
}

\lref\BrignoleCM{
  A.~Brignole, J.~A.~Casas, J.~R.~Espinosa and I.~Navarro,
  ``Low-scale supersymmetry breaking: Effective description, electroweak
  breaking and phenomenology,''
  Nucl.\ Phys.\  B {\bf 666}, 105 (2003)
  [arXiv:hep-ph/0301121].
  }

\lref\AffleckXZ{
  I.~Affleck, M.~Dine and N.~Seiberg,
  ``Dynamical Supersymmetry Breaking In Four-Dimensions And Its
  Phenomenological Implications,''
  Nucl.\ Phys.\  B {\bf 256}, 557 (1985).
}

\lref\DineYW{
  M.~Dine and A.~E.~Nelson,
  ``Dynamical supersymmetry breaking at low-energies,''
  Phys.\ Rev.\  D {\bf 48}, 1277 (1993)
  [arXiv:hep-ph/9303230].
}

\lref\DineVC{
  M.~Dine, A.~E.~Nelson and Y.~Shirman,
  ``Low-Energy Dynamical Supersymmetry Breaking Simplified,''
  Phys.\ Rev.\  D {\bf 51}, 1362 (1995)
  [arXiv:hep-ph/9408384].
}

\lref\LutyFK{
  M.~A.~Luty,
  ``Naive dimensional analysis and supersymmetry,''
  Phys.\ Rev.\  D {\bf 57}, 1531 (1998)
  [arXiv:hep-ph/9706235].
}

\lref\DineAG{
  M.~Dine, A.~E.~Nelson, Y.~Nir and Y.~Shirman,
  ``New tools for low-energy dynamical supersymmetry breaking,''
  Phys.\ Rev.\  D {\bf 53}, 2658 (1996)
  [arXiv:hep-ph/9507378].
}

\lref\WittenKV{
  E.~Witten,
  ``Mass Hierarchies In Supersymmetric Theories,''
  Phys.\ Lett.\  B {\bf 105}, 267 (1981).
}

\lref\DineQJ{
  M.~Dine,
  ``Some issues in gauge mediation,''
  Nucl.\ Phys.\ Proc.\ Suppl.\  {\bf 62}, 276 (1998)
  [arXiv:hep-ph/9707413].
}

\lref\CohenVB{
  A.~G.~Cohen, D.~B.~Kaplan and A.~E.~Nelson,
  ``The more minimal supersymmetric standard model,''
  Phys.\ Lett.\  B {\bf 388}, 588 (1996)
  [arXiv:hep-ph/9607394].
}

\lref\BanksMG{
  T.~Banks and V.~Kaplunovsky,
  ``Nosonomy Of An Upside Down Hierarchy Model. 1,''
  Nucl.\ Phys.\  B {\bf 211}, 529 (1983).
}
\lref\KaplunovskyYX{
  V.~Kaplunovsky,
  ``Nosonomy Of An Upside Down Hierarchy Model. 2,''
  Nucl.\ Phys.\  B {\bf 233}, 336 (1984).
}

\lref\DimopoulosGM{
  S.~Dimopoulos and S.~Raby,
  ``Geometric Hierarchy,''
  Nucl.\ Phys.\  B {\bf 219}, 479 (1983).
}

\lref\DermisekQJ{
  R.~Dermisek, H.~D.~Kim and I.~W.~Kim,
  ``Mediation of supersymmetry breaking in gauge messenger models,''
  JHEP {\bf 0610}, 001 (2006)
  [arXiv:hep-ph/0607169].
}

\lref\DineGU{
  M.~Dine and W.~Fischler,
  ``A Phenomenological Model Of Particle Physics Based On Supersymmetry,''
  Phys.\ Lett.\  B {\bf 110}, 227 (1982).
}

\lref\NappiHM{
  C.~R.~Nappi and B.~A.~Ovrut,
  ``Supersymmetric Extension Of The SU(3) X SU(2) X U(1) Model,''
  Phys.\ Lett.\  B {\bf 113}, 175 (1982).
}

\lref\DineZB{
  M.~Dine and W.~Fischler,
  ``A Supersymmetric Gut,''
  Nucl.\ Phys.\  B {\bf 204}, 346 (1982).
}

\lref\AlvarezGaumeWY{
  L.~Alvarez-Gaume, M.~Claudson and M.~B.~Wise,
  ``Low-Energy Supersymmetry,''
  Nucl.\ Phys.\  B {\bf 207}, 96 (1982).
}

\lref\tobenomura{
  Y.~Nomura and K.~Tobe,
  ``Phenomenological aspects of a direct-transmission model of dynamical
  supersymmetry breaking with the gravitino mass m(3/2) $<$ 1-keV,''
  Phys.\ Rev.\  D {\bf 58}, 055002 (1998)
  [arXiv:hep-ph/9708377].
}

\lref\IzawaGS{
  K.~I.~Izawa, Y.~Nomura, K.~Tobe and T.~Yanagida,
  ``Direct-transmission models of dynamical supersymmetry breaking,''
  Phys.\ Rev.\  D {\bf 56}, 2886 (1997)
  [arXiv:hep-ph/9705228].
}

\lref\CheungES{
  C.~Cheung, A.~L.~Fitzpatrick and D.~Shih,
  ``(Extra)Ordinary Gauge Mediation,''
  arXiv:0710.3585 [hep-ph].
}

\lref\hiddenren{
  A.~G.~Cohen, T.~S.~Roy and M.~Schmaltz,
  JHEP {\bf 0702}, 027 (2007)
  [arXiv:hep-ph/0612100].
}

\lref\dimgiud{
 S.~Dimopoulos and G.~F.~Giudice,
  ``Multi-messenger theories of gauge-mediated supersymmetry breaking,''
  Phys.\ Lett.\  B {\bf 393}, 72 (1997)
  [arXiv:hep-ph/9609344].
}

\lref\pierre{S.~P.~Martin and P.~Ramond,
  ``Sparticle spectrum constraints,''
  Phys.\ Rev.\  D {\bf 48}, 5365 (1993)
  [arXiv:hep-ph/9306314].
}

\lref\faraggi{
  A.~E.~Faraggi, J.~S.~Hagelin, S.~Kelley and D.~V.~Nanopoulos,
  ``Sparticle Spectroscopy,''
  Phys.\ Rev.\  D {\bf 45}, 3272 (1992).
}

\lref\kawamura{
  Y.~Kawamura, H.~Murayama and M.~Yamaguchi,
  ``Probing symmetry breaking pattern using sfermion masses,''
  Phys.\ Lett.\  B {\bf 324}, 52 (1994)
  [arXiv:hep-ph/9402254].
}

\lref\MartinZB{
  S.~P.~Martin,
  ``Generalized messengers of supersymmetry breaking and the sparticle mass
  spectrum,''
  Phys.\ Rev.\  D {\bf 55}, 3177 (1997)
  [arXiv:hep-ph/9608224].
}

\lref\spectroscopy{
 S.~Dimopoulos, S.~D.~Thomas and J.~D.~Wells,
  ``Sparticle spectroscopy and electroweak symmetry breaking with
  gauge-mediated supersymmetry breaking,''
  Nucl.\ Phys.\  B {\bf 488}, 39 (1997)
  [arXiv:hep-ph/9609434].
}

\lref\polchinski{ J.~Polchinski and L.~Susskind, ``Breaking Of
Supersymmetry At Intermediate-Energy,''
  Phys.\ Rev.\  D {\bf 26}, 3661 (1982).
 }

\lref\dgp{ G.~R.~Dvali, G.~F.~Giudice and A.~Pomarol,
  ``The $\mu$-Problem in Theories with Gauge-Mediated Supersymmetry Breaking,''
  Nucl.\ Phys.\  B {\bf 478}, 31 (1996)
  [arXiv:hep-ph/9603238].
}

\lref\martinmu{ T.~S.~Roy and M.~Schmaltz, ``A hidden solution to
the $\mu/B_\mu$ problem in gauge mediation,''
  arXiv:0708.3593 [hep-ph].
}

\lref\hidrentwo{
  H.~Murayama, Y.~Nomura and D.~Poland,
  ``More Visible Effects of the Hidden Sector,''
  arXiv:0709.0775 [hep-ph].
}

\lref\DineXI{
  M.~Dine, N.~Seiberg and S.~Thomas,
  ``Higgs Physics as a Window Beyond the MSSM (BMSSM),''
  Phys.\ Rev.\  D {\bf 76}, 095004 (2007)
  [arXiv:0707.0005 [hep-ph]].
}

\lref\WessCP{
  J.~Wess and J.~Bagger,
  ``Supersymmetry and supergravity,''
{\it  Princeton, USA: Univ. Pr. (1992) 259 p}
}

\lref\MartinNS{
  S.~P.~Martin,
  ``A supersymmetry primer,''
  arXiv:hep-ph/9709356.
}

\lref\WittenNF{
  E.~Witten,
  ``Dynamical Breaking Of Supersymmetry,''
  Nucl.\ Phys.\  B {\bf 188}, 513 (1981).
}

\lref\IzawaPK{
  K.~I.~Izawa and T.~Yanagida,
  ``Dynamical Supersymmetry Breaking in Vector-like Gauge Theories,''
  Prog.\ Theor.\ Phys.\  {\bf 95}, 829 (1996)
  [arXiv:hep-th/9602180].
}

\lref\IntriligatorPU{
  K.~A.~Intriligator and S.~D.~Thomas,
 ``Dynamical Supersymmetry Breaking on Quantum Moduli Spaces,''
  Nucl.\ Phys.\  B {\bf 473}, 121 (1996)
  [arXiv:hep-th/9603158].
}

\lref\RoyNZ{
  T.~S.~Roy and M.~Schmaltz,
  ``A hidden solution to the mu/Bmu problem in gauge mediation,''
  Phys.\ Rev.\  D {\bf 77}, 095008 (2008)
  [arXiv:0708.3593 [hep-ph]].
}

\lref\MurayamaGE{
  H.~Murayama, Y.~Nomura and D.~Poland,
  ``More Visible Effects of the Hidden Sector,''
  Phys.\ Rev.\  D {\bf 77}, 015005 (2008)
  [arXiv:0709.0775 [hep-ph]].
}

\lref\CheungES{
  C.~Cheung, A.~L.~Fitzpatrick and D.~Shih,
  ``(Extra)Ordinary Gauge Mediation,''
  JHEP {\bf 0807}, 054 (2008)
  [arXiv:0710.3585 [hep-ph]].
}

\lref\GrisaruVE{
  M.~T.~Grisaru, M.~Rocek and R.~von Unge,
  ``Effective K\"ahler Potentials,''
  Phys.\ Lett.\  B {\bf 383}, 415 (1996)
  [arXiv:hep-th/9605149].
}

\lref\IntriligatorDD{
  K.~A.~Intriligator, N.~Seiberg and D.~Shih,
  ``Dynamical SUSY breaking in meta-stable vacua,''
  JHEP {\bf 0604}, 021 (2006)
  [arXiv:hep-th/0602239].
}

\lref\PerezNG{
  G.~Perez, T.~S.~Roy and M.~Schmaltz,
  ``Phenomenology of SUSY with scalar sequestering,''
  arXiv:0811.3206 [hep-ph].
}

\lref\GirardelloWZ{
  L.~Girardello and M.~T.~Grisaru,
  ``Soft Breaking Of Supersymmetry,''
  Nucl.\ Phys.\  B {\bf 194}, 65 (1982).
}

\lref\MeadeWD{
  P.~Meade, N.~Seiberg and D.~Shih,
  ``General Gauge Mediation,''
  arXiv:0801.3278 [hep-ph].
}

\lref\EGGM{
  M.~Buican, P.~Meade, N.~Seiberg and D.~Shih,
  ``Exploring General Gauge Mediation,''
 to appear.
}

\lref\YanagidaYF{
  T.~Yanagida,
  ``A solution to the mu problem in gauge-mediated supersymmetry-breaking
  models,''
  Phys.\ Lett.\  B {\bf 400}, 109 (1997)
  [arXiv:hep-ph/9701394].
}

\lref\DimopoulosJE{
  S.~Dimopoulos, G.~R.~Dvali and R.~Rattazzi,
  ``A simple complete model of gauge-mediated SUSY-breaking and dynamical
  relaxation mechanism for solving the mu problem,''
  Phys.\ Lett.\  B {\bf 413}, 336 (1997)
  [arXiv:hep-ph/9707537].
}

\lref\LangackerHS{
  P.~Langacker, N.~Polonsky and J.~Wang,
  ``A low-energy solution to the mu-problem in gauge mediation,''
  Phys.\ Rev.\  D {\bf 60}, 115005 (1999)
  [arXiv:hep-ph/9905252].
}

\lref\HallUP{
  L.~J.~Hall, Y.~Nomura and A.~Pierce,
  ``R symmetry and the mu problem,''
  Phys.\ Lett.\  B {\bf 538}, 359 (2002)
  [arXiv:hep-ph/0204062].
}

\lref\LiuPA{
  T.~Liu and C.~E.~M.~Wagner,
  ``Dynamically Solving the $\mu/B_\mu$ Problem in Gauge-mediated Supersymmetry
  Breaking,''
  JHEP {\bf 0806}, 073 (2008)
  [arXiv:0803.2895 [hep-ph]].
}

\lref\BatraRC{
  P.~Batra and E.~Ponton,
  ``The Supersymmetric Higgs,''
  arXiv:0809.3453 [hep-ph].
}

\Title{\vbox{\baselineskip12pt }} {\vbox{\centerline{
$\mu$ and General Gauge Mediation}}}
\smallskip
\centerline{Zohar Komargodski and Nathan
Seiberg}
\smallskip
\bigskip
\centerline{{\it School
of Natural Sciences, Institute for Advanced Study, Princeton, NJ
08540 USA}} \vskip 1cm

\noindent
We address the $\mu$-problem in the context of General Gauge Mediation (GGM). We classify possible models depending on the way the Higgs fields couple to the supersymmetry breaking hidden-sector.  The different types of models have distinct signatures in the MSSM parameters.  We find concrete and surprisingly simple examples based on messengers in each class.  These examples lead to all the soft masses and a consistent Higgs-sector.

\Date{December 2008}

\newsec{Introduction and Summary}

Supersymmetry is arguably the leading candidate for TeV physics to be discovered at the LHC.  An important open question is the identification of the supersymmetry breaking mechanism and its scale.  One possibility is that supersymmetry is broken by a ``hidden-sector" at low energy and the information about supersymmetry breaking is communicated to the standard model fields via gauge interactions~\refs{\DineGU\NappiHM\DineZB\AlvarezGaumeWY\DineYW\DineVC-\DineAG}.  These models naturally solve the supersymmetric flavor problem.  Recently the authors of~\MeadeWD\ have defined the most general model of this kind by the criterion that as the three MSSM gauge couplings $\alpha_{i=1,2,3}$ are set to zero, the MSSM is decoupled from the hidden-sector.  Despite this weak assumption, the framework of General Gauge Mediation (GGM) leads to specific predictions about the MSSM parameters~\refs{\MeadeWD,\EGGM}.  Using data from future experiments we will be able to examine these predictions and to learn whether the idea of gauge mediation is correct or not.

However, it is well known that gauge mediation does not address the $\mu$-problem.  Here $\mu$ is the coefficient in the MSSM coupling
 \eqn\mudef{\int d^2 \theta \mu \CH_u\CH_d~,}
where we denote the superfield by $\CH$ and its bottom component by $H$.

The $\mu$-problem has several aspects.  First, one would like to explain why $\mu$ is of the same order of magnitude as the various soft supersymmetry breaking terms in the MSSM:
 \eqn\musoft{\mu^2 \sim m_{soft}^2~.}
It is clear that in order to solve this problem in the context of gauge mediation, we need to relax the definition given above.  In particular, the Higgs-sector of the MSSM must be directly coupled to the hidden-sector.
Then, $\mu$ is related to the hidden-sector scale $M$, which can be dynamically generated, thus making it naturally small.  Once $\mu$ is set to the right order of magnitude, the supersymmetric nonrenormalization theorem guarantees that it remains of that magnitude.

The second aspect of the $\mu$-problem makes it much more difficult to solve, especially in the context of gauge mediation.  The Higgs-sector includes various supersymmetry breaking terms.  There are quadratic terms
 \eqn\quaddef{m_u^2 |H_u|^2 + m_d^2 |H_d|^2 +(B_\mu H_uH_d + c.c.)~,}
cubic $A$-terms \eqn\Atermsdef{A^u_{ij} H_u Q^i \bar u^j + A^d_{ij} H_d Q^i \bar d^j + A^L_{ij} H_d L^i \bar E^j}
(here and below we use the standard notation for the MSSM matter fields), ``wrong Higgs couplings"\foot{These terms are rarely studied because they have high effective dimension and therefore are not included in the MSSM. (For a discussion of effective dimensions in the MSSM, see e.g.\ \DineXI.)}
 \eqn\wrongHdef{A'^u_{ij} H_d^\dagger Q^i \bar u^j + A'^d_{ij} H_u^\dagger Q^i \bar d^j + A'^L_{ij} H_u^\dagger L^i \bar E^j~,}
and various higher order terms.  As we will review below, a typical coupling of the Higgs fields to the hidden-sector leads to the unacceptable relations: $\mu^2 \ll B_\mu, m_{u,d}^2$.  So for electroweak symmetry breaking the challenge is to ensure
 \eqn\mustensure{\mu^2 \sim B_\mu \sim m_{u,d}^2 \sim m_{soft}^2~.}
Because of this challenge people often refer to this problem as the $\mu/B_\mu$-problem.

The soft terms $m_{u,d}^2$ receive contributions both from the gauge mediation mechanism and from the additional new couplings of the Higgs fields to the hidden-sector.  To leading order the gauge mediation contributions to $m_u^2$ and $m_d^2$ are the same and they are equal to that of the slepton doublet $m_{\tilde L}^2$.  The new  contributions which will be discussed below will be denoted by $\delta m_{u,d}^2$. Similarly, we write for the $A$-terms $\delta a_{u,d}$ when referring to the contributions induced by the new couplings.

We will find it convenient to follow~\MeadeWD\ and to classify the models by the $SU(2)$ representation of the hidden-sector operators which couple to the Higgs fields.  The first kind of models involve $SU(2)$ doublet hidden-sector operators $\Phi_{u,d}$ which couple to the Higgs fields through the superpotential coupling
 \eqn\doubcoup{\int d^2 \theta \left( \lambda_u \CH_u \Phi_d + \lambda_d \CH_d \Phi_u\right)~.}
These models will be discussed in section 2.  The second class of models involve an $SU(2)$ singlet hidden-sector operator $\CS$, which couples to the Higgs fields as
 \eqn\singcoup{\int d^2 \theta \lambda^2 \CS \CH_u \CH_d~.}
($\CS$ is the superfield and its bottom component is $S$.)  The definition of the coupling constant here as $\lambda^2$ is for easy comparison with the class \doubcoup.  These models will be discussed in section 3.

We should emphasize that $\Phi_{u,d}$ in \doubcoup\ and $\CS$ in \singcoup\ are not necessarily elementary fields -- we have in mind a situation where these could be composite operators in a complicated quantum field theory.  Specific examples of these two classes have been studied in the literature.  For example, models of the kind \doubcoup\ with $\Phi_{u,d}$ bilinear of other fields were discussed in \refs{\DvaliCU\MurayamaGE-\CsakiSR} and models like \singcoup\ with $\CS$ an elementary fields are known as NMSSM  (see e.g.\ \refs{\DvaliCU,\DineXK,\GiudiceCA}). For additional examples and other approaches to the problem, see e.g.\ \refs{\YanagidaYF\DimopoulosJE\LangackerHS\HallUP-\LiuPA} and references therein.

We will further divide each class of models into two subclasses depending on the scales in the problem.  First, the hidden-sector can be characterized by a single scale $M$ and the scale of supersymmetry breaking is also given by $M$. As we will see, these models cannot satisfy our condition \mustensure\ with perturbative $\lambda$.  In the second subclass the hidden-sector is characterized by two mass scales.  There is some dynamics at the scale $M$ but supersymmetry is broken at a lower energy with
 \eqn\FMrel{F \ll M^2~.}
Such models turn out to be more promising.

The two classes of models \doubcoup\ and \singcoup\ have some common predictions.  The coefficients of the $A$-terms are small and are determined by the couplings
 \eqn\adef{a_uF_{H_u}^\dagger H_u+ a_d F_{H_d}^\dagger H_d~.}
The ``wrong-Higgs couplings" \wrongHdef\ are even smaller and are determined by
 \eqn\wrongHd{a'_u F_{H_d} H_u+a'_d F_{H_u} H_d~.}
After integrating out the auxiliary fields $F_{H_{u,d}}$ in the presence of the operators \adef,\wrongHd\ we find the terms \Atermsdef,\wrongHdef\ which are proportional to the Yukawa couplings, and hence satisfy the minimal flavor violation assumption.

In addition, our two classes of models have some different signatures.  Models with doublets satisfy the relations \mustensure, while the models with a singlet have a more specific prediction
 \eqn\singsig{m_u^2 = m_d^2 = m_{\tilde L}^2~,}
where again, $m_{\tilde L}^2$ is the left handed slepton mass squared.  This relation represents the gauge mediation contribution without additional contribution from the coupling \singcoup.  It is evaluated at the scale $M$ and must be renormalized down.

Therefore, once the soft terms are measured, they will be able to tell us whether the microscopic model is of the gauge mediation type and will also point to the class of models of the Higgs coupling.  We also note that in both of our models the coefficients of the effective dimension five operators of \DineXI\ turn out to be very small.

In section 4 we will present specific models in our two classes.  They will demonstrate that these classes indeed include viable models.  Our examples are simpler than the known models in the literature, thus opening the way to a more detailed model building work.

Finally, we would like to make one more comment.  In this paper we follow the line of thought of general gauge mediation \MeadeWD\ assuming that all the MSSM fields including the Higgs fields are elementary.  However, some of the gauge mediation results and some of the results here are also valid when some of these fields are composite. Proposals in which certain MSSM fields are composite include, for instance \refs{\CohenVB,\DineQJ}.

\newsec{Coupling the Hidden-Sector to the Higgs Fields via $SU(2)$ Doublets}

In this section we suppose that the gauge symmetry of the MSSM, $SU(3)\times SU(2)\times U(1)$, is embedded in a flavor symmetry of the hidden-sector, and consider two $SU(2)$ doublet operators of the hidden-sector $\Phi_{u}$, $\Phi_d$ with a superpotential coupling them to the Higgs superfields
\eqn\supgen{\CW=\lambda_u\CH_u\Phi_d+\lambda_d\CH_d\Phi_u~.}
We think of $\Phi_{u,d}$ as having dimension 2 and $\lambda_{u,d}$ are dimensionless.
The bare $\mu$ term does not appear in the superpotential. This could be due to a symmetry.

Integrating out the hidden sector dynamics leads to various terms in the effective theory.  The quadratic terms in the Higgs-sector at zero momentum are
\eqn\Lagr{{\cal L}={\cal Z}_uF_{H_u}F_{H_u}^\dagger+{\cal Z}_dF_{H_d}F_{H_d}^\dagger+\int d^2\theta\left(\CW_{MSSM}^{\ \mu=0}+\mu \CH_u\CH_d\right) +{\cal L}_{soft}+{\cal L}_{hard}~,}
with
\eqn\lagsoft{\eqalign{
&-{\cal L}_{soft}= m^2_{u}H_uH_u^\dagger+m^2_{d}H_dH_d^\dagger+\left(B_\mu H_uH_d+c.c.\right)+
\left( a_uH_uF_{H_u}^\dagger+a_dH_dF_{H_d}^\dagger+c.c.\right)
~,\cr &-{\cal L}_{hard}= \left( a'_uH_uF_{H_d}+a'_dH_dF_{H_u}+c.c.\right)+\left(\gamma F_{H_u}F_{H_d}+c.c.\right)~.}}

A few comments are in order. The hard term $\gamma$ gives rise to an interaction of four MSSM sparticles once the auxiliary fields are eliminated. It is of large effective dimension and usually neglected, here we present it for the completeness of the Lagrangian to quadratic order.
Secondly, it is important to stress that the $\mu,m^2_{{u,d}},B_\mu$ terms of \Lagr,\lagsoft\ are not the physically measurable values as we have not written corrections to the wave functions of the Higgs and Higgsino particles. These corrections appear in the full momentum-dependent effective theory. (On the other hand, corrections to the normalization of $|F_{H_{u,d}}|^2$ appear already at zero momentum.) If $\lambda_{u,d}$ are small parameters these momentum dependent corrections are negligible and the physical values can be safely read from the zero momentum effective theory.

\subsec{Correlation Functions and their Properties}
The Higgs parameters appearing in \Lagr,\lagsoft\ are all calculable in terms of (zero momentum) correlation functions of the hidden-sector. The complete set of the relevant  two-point functions of the hidden-sector is
\eqn\twopoint{\matrix{ R_{0,0}^{u,d}(p)=\langle \phi_{u,d}\phi_{u,d}^\dagger\rangle(p)~,\qquad & F_{0,0}(p)=\langle \phi_{u}\phi_{d}\rangle(p)~, \cr
R_{1,1}^{u,d}(p)=\langle \psi_{\phi_{u,d}}\bar\psi_{\phi_{u,d}}\rangle(p) ~,\qquad & F_{1,1}(p)=\langle \psi_{\phi_{u}}\psi_{\phi_{d}}\rangle(p) ~, \cr
R_{2,0}^{u,d}(p)=\langle F_{\phi_{u,d}}\phi_{u,d}^\dagger\rangle(p) ~,\qquad & F_{2,0}(p)=\langle F_{\phi_{u}}\phi_{d}\rangle(p) ~,\cr
R_{0,2}^{u,d}(p)=(R_{2,0}^{u,d}(p))^\dagger~,\qquad  & F_{0,2}(p)=\langle \phi_{u}F_{\phi_{d}}\rangle(p)~,  \cr
R_{2,2}^{u,d}(p)=\langle F_{\phi_{u,d}}F_{\phi_{u,d}}^\dagger\rangle(p)~,\qquad  & F_{2,2}(p)=\langle F_{\phi_{u}}F_{\phi_{d}}\rangle(p)~, }}
where $\phi_{u,d}$ are the bottom components of $\Phi_{u,d}$.  Some of these correlation functions can be non-zero in a SUSY theory and the rest arise only due to SUSY breaking. To analyze this more precisely we write some combinations of these correlation functions as variations under supersymmetry transformations of  other two-point functions. First, there are Ward-identities for the $R_{ij}$ correlation functions of \twopoint\ (which follow from $[\bar Q,\phi]=0$)
\eqn\wardidentonep{\eqalign{
&{1\over 2}\langle \{Q_\alpha, [\bar Q_{\dot \alpha},\phi_{u,d}\phi_{u,d}^\dagger]\}\rangle(p)= \left(R_{1,1}^{u,d}(p)\right)_{\alpha\dot\alpha}-p_\mu\sigma^\mu_{\alpha\dot\alpha}
R_{0,0}^{u,d}(p) ~,\cr &{1\over 4}\langle \{Q^\alpha,[Q_\alpha,\phi_{u,d}\phi_{u,d}^\dagger ]\}\rangle(p)=R_{2,0}^{u,d}(p) ~,\cr
&{1\over 16}\langle \{ Q^\alpha,[ Q_\alpha,\{\bar Q_{\dot\alpha}, [\bar Q^{\dot\alpha},\phi_{u,d}\phi_{u,d}^\dagger]\}]\} \rangle(p)=R_{2,2}^{u,d}(p)-p_\mu\sigma^{\mu}_{\alpha\dot\beta}\left(R_{1,1}
^{u,d}(p)\right)^{\alpha\dot\beta}-p^2R_{0,0}^{u,d}(p)~.}}
There are also similar identities for the $F_{ij} $ correlation functions of \twopoint,
\eqn\wardFonep{\eqalign{ &{1\over 4}\langle\{\bar Q_{\dot\alpha},[\bar Q^{\dot\alpha} ,F_{\phi_u}\phi_d]\}\rangle(p)=-p^2F_{0,0}(p)~,\cr
 & {1\over2} \langle\{ Q_{\alpha},[\bar Q_{\dot\alpha} ,F_{\phi_u}\phi_d]\}\rangle(p)=p_\mu\sigma^\mu_{\lambda\dot\alpha}\left(\delta^
 \lambda_\alpha  F_{2,0}(p)- \left(F_{1,1}(p)\right)^{\lambda}_{\ \ \alpha}\right) ~, \cr
 & {1\over2}\langle \{ Q_{\alpha},[\bar Q_{\dot\alpha} ,\phi_uF_{\phi_d}]\}\rangle(p)=-p_\mu\sigma^\mu_{\lambda\dot\alpha}\left(\delta^
 \lambda_\alpha  F_{0,2}(p)+ \left(F_{1,1}(p)\right)^{\ \ \lambda}_{\alpha} \right) ~,\cr
& {1\over 4}\langle\{ Q^{\alpha},[ Q_{\alpha} ,\phi_uF_{\phi_d}]\}\rangle(p)= F_{2,2}(p)~.}}

Obviously all the combinations of correlation functions on the right hand side of {\wardidentonep-\wardFonep} vanish if SUSY is unbroken. Suppose SUSY is broken and the vacuum energy density is $F^2$. Further, let us assume that this scale is smaller than the typical mass scale in the problem $M$ and then all the observables can be expanded in a power series in $F/M^2$.  We will not describe it here, but a careful analysis shows that in this limit correlation functions like $\langle \{Q,[Q,(\cdot)]\}\rangle $ can begin at linear order in $F$ while correlation functions like $\langle \{Q,[Q^\dagger,(\cdot)]\}\rangle $ necessarily begin at least at order $F^2$. (We will provide some consistency checks though.) Of course, symmetries or some other considerations can make both correlation functions begin at an even higher order in $F$ and below we will see some examples of this phenomenon.

In terms of the correlation functions \twopoint\ we can evaluate the terms presented in \Lagr,\lagsoft. To leading order in the
$\lambda_u,\lambda_d$ couplings the dictionary is
\eqn\dict{\eqalign{&
\mu={i\over 2}\lambda_u\lambda_dF_{1,1}(p=0)~,\cr
& \delta m^2_{u,d}=-i|\lambda_{u,d}|^2 R_{2,2}^{d,u} (p=0) ~,\cr
 &B_\mu=-i\lambda_u\lambda_dF_{2,2}(p=0) ~,\cr
 &\delta a_{u,d}=-i|\lambda_{u,d}|^2R_{2,0}^{d,u}(p=0)~,\cr
  &a'_u=-i\lambda_u\lambda_d\left(F_{2,0}(p=0)-{1\over 2}F_{1,1}(p=0)\right)~,\cr
 &a'_d=-i\lambda_u\lambda_d\left(F_{0,2}(p=0)-{1\over 2}F_{1,1}(p=0)\right)~,\cr
 &\gamma=-i\lambda_u\lambda_dF_{0,0}(p=0)~,\cr
 & \delta{\cal Z}_{u,d}=i|\lambda_{u,d}|^2R_{0,0}^{d,u}(p=0)
~.}}

All of these correlation functions except the one defining $\delta{\cal Z}_{u,d}$ can be shown to be well defined at zero momentum in the sense of being UV insensitive. This is of course what we expect to find in a theory with spontaneously broken supersymmetry in the hidden-sector. However, the function $R_{0,0}^{u,d}(p=0)$ generally has a logarithm of the UV scale. The reason is that this is precisely the supersymmetric wave function renormalization. As we have remarked below \lagsoft, in a systematic expansion in $\lambda_{u,d}$ wave function renormalization plays no role for the values of physical low-energy observables and hence this UV sensitivity is irrelevant for our purposes.

There is a nice way to write all of these soft terms using an effective action packaged in superspace. First let us discuss the terms proportional to either of $|\lambda_u^2|$, $|\lambda_d|^2$. These two-point function contributions to the MSSM soft terms can be summarized in an effective action for the Higgs fields of the form
\eqn\effnonh{S_{eff}=i\int {d^4p\over (2\pi)^4}\int d^4\theta\left( |\lambda_u|^2\CH_u\CH_u^\dagger \langle\Phi_d(p) \Phi_d^\dagger(-p) \rangle+ |\lambda_d|^2\CH_d\CH_d^\dagger \langle\Phi_u(p) \Phi_u^\dagger(-p) \rangle\right)~, }
where $\Phi_{u,d}$ are superfields. There is another way to write it which emphasizes the order in $F$ at which each term begins
\eqn\effnonhexp{S_{eff}=i\int {d^4p\over (2\pi)^4}\int d^4\theta \left(|\lambda_u|^2\CH_u\CH_u^\dagger \langle e^{i\left(\theta Q+\bar\theta\bar Q\right)}\phi_d(p) \phi_d^\dagger(-p)e^{-i\left(\theta Q+\bar\theta\bar Q\right)} \rangle+ (u\leftrightarrow d)\right)~.}
The different observables mentioned in \dict\ arise by expanding \effnonhexp\ in components and using the Ward-identities \wardidentonep. The bottom component correlation function $\langle\phi\phi^\dagger\rangle$ is the supersymmetric wave function renormalization. The $\theta\bar\theta$ represents corrections from the non-supersymmetric dynamics to the wave function of the fermion. This correction arises at order $F^2$. The $\theta^2,\bar\theta^2$ terms are the $A$-terms, and indeed they arise at order $F$. Lastly, the $\theta^2\bar\theta^2$ is responsible for the soft masses $\delta m^2_{{u,d}}$ as well as non-supersymmetric corrections to the wave function of the Higgs boson. Both of these effects are at order $F^2$ or higher.

Now, we turn to the superspace effective action for the terms proportional to $\lambda_u\lambda_d$. The result here is
\eqn\effkah{S'_{eff}=i\lambda_u\lambda_d\int {d^4p\over (2\pi)^4}\int d^4\theta {1\over 16p^2}\CH_u(p)D^2\CH_d(-p)\left\langle D^2\Phi_u(p)\Phi_d(-p)\right\rangle+c.c.~.}
We can write, analogously to \effnonhexp,
\eqn\expo{{1\over 4}\left\langle D^2\Phi_u(p)\Phi_d(-p)\right\rangle=\left\langle e^{i\left(\theta Q+\bar\theta\bar Q\right)}F_{\phi_u}(p)\phi_d(-p)e^{-i\left(\theta Q+\bar\theta\bar Q\right)}\right\rangle~,}
from which we see directly that the bottom component contribution exists already in the SUSY limit\foot{The bottom component is just $\langle F_{\phi_u}\phi_d\rangle$ which is related by SUSY transformations to a correlation function of fermions and hence can arise in the SUSY limit. This is the content of the second and third equations in \wardFonep.} and gives rise to the $\mu$ term, while the others arise only once SUSY is broken. The $B_\mu$ term arises from the $\theta^2$ component of \expo\ (and, therefore, may begin at order $F$). The $\bar\theta^2$ term in \expo\ gives rise to non-supersymmetric contributions to the two-point function $F_{H_u}F_{H_d}$ for the Higgs fields (this eventually generates interactions among four MSSM sparticles and higher order corrections to $B_\mu$ and the $A$-terms). This effect also arises at order $F$ in general. The $\theta\bar\theta$ , $\theta^2\bar\theta^2$ components provide the non-supersymmetric $F^2$ corrections to the correlation functions $H_uF_{H_d} , H_dF_{H_u}$ which give rise to the wrong-Higgs couplings. It is important that although there is a $p^2$ in the denominator in \effkah\ the resulting component Lagrangian is regular.

Another, complementary, organizing principle for \dict\ is to write the effective operators which can generate them at low energies using a SUSY breaking spurion field. This is also a way to make simple consistency checks on the order in SUSY breaking at which these observables are generated. We, therefore, write all the terms with a spurion
\eqn\spurion{\CX=\theta^2 {F \over M}~.}
In this convention the spurion is dimensionless and the terms which are strictly soft do not contain any further suppression by a high energy scale $M$.  The relevant terms are
\eqn\supspu{{\cal W}= \mu \CH_u\CH_d+M\CH_u\CH_d\CX~,}
\eqn\kahlerspu{{\cal K}=\CH_u\CH_u^\dagger \left(\CX+\CX^\dagger+\CX\CX^\dagger\right)+\CH_d\CH_d^\dagger \left(\CX+\CX^\dagger+\CX\CX^\dagger\right)~.}
The coefficients of all these operators are in general independent. There are two operators we have not written down in the K\"ahler potential, $\CH_u\CH_d\CX^\dagger$, $\CH_u\CH_d\CX\CX^\dagger$. The reason is that these only shift the $\mu$ and $B_\mu$ terms, so these operator can be absorbed in the superpotential.\foot{Another possible basis is to redefine holomorphically the Higgs superfield $\CH=\CH(1+\CX)$ with the appropriate coefficient. This allows one to eliminate the linear terms in $\CX$ in \kahlerspu, but reintroduces new terms in the MSSM superpotential coupling $\CX$ directly to the quarks and leptons. Hence, for simplicity we prefer to use this basis.}

It is easy to see that these operators give rise to all the soft terms. In addition, we see that the $A$-terms and the $B_\mu$ term arise at linear order in $F$ while the $\delta m^2_{{u,d}}$ parameters arise only at the order $F^2$. In our language the latter property follows from the quartic commutator in \wardidentonep, $\langle FF^\dagger \rangle(p=0) =
\langle \{Q, [Q,\{\bar Q,[\bar Q, (\cdot) ]\}]\}\rangle $. The familiar expressions \supspu,\kahlerspu\ are valid to leading order in $F$. Our general expressions \effnonhexp,\effkah\ generalize them to all orders.

As we have already discussed, there are also some other induced contributions at low energies which are hard. For instance, the operator in the K\"ahler potential
\eqn\nonsoft{\delta{\cal K}={1\over M}\CH_u\left(D^2 \CH_d\right)\CX\CX^\dagger}
generates the operator $ H_uF_{H_d}$ which gives rise to the wrong-Higgs couplings $a'$. Indeed, as we expect from \wardFonep\ these can arise only at order $F^2/M^3$. The fact that we need covariant derivatives to generate this term fits nicely with the superspace description \effkah\ where we had to use covariant derivatives to write the superspace effective action.

\subsec{One-Scale Models}
First, let us assume that the hidden-sector is a one-scale model, with scale $M$. It is easy to determine all the soft terms by dimensional analysis. To leading order in $\lambda_{u,d}$ we have  \eqn\onescale{\eqalign{& \mu\sim \lambda_u\lambda_d M~,\cr & B_\mu\sim \lambda_u\lambda_d M^2~,\cr &\delta m^2_{{u,d}}\sim \left|\lambda_{u,d}\right|^2M^2~, \cr & \delta a_{u,d}\sim \left|\lambda_{u,d}\right|^2M~.}}
This results in
\eqn\constr{B_\mu^2\sim \delta m^2_{u}\delta m^2_{d}\sim \mu^2 M^2~.}
We assume $\lambda_{u,d}\ll 1$ so that we can expand in $\lambda_{u,d}$ and ignore backreaction on the hidden-sector. Therefore,  $M\gg \mu$ (this also follows from the formula for the gaugino masses $m_{\tilde g}\sim \alpha_g M\ll M$) resulting in
$B_\mu,\delta m^2_{{u,d}}\gg \mu^2$ which is problematic for viable electroweak symmetry breaking.

We conclude that this scenario is inconsistent.  A possible way out is to consider parameterically strong mixing between the Higgs fields and the hidden-sector \CsakiSR.  However, this parameterically strong coupling makes the analysis challenging and more work is needed in order to establish the viability of this approach.

It is important to point out here, that from this vantage point, theories in which there is a scale $M$ and some (usually different) SUSY breaking scale $F$ are actually effectively one-scale models if all the soft terms and the $\mu$ term are induced by SUSY breaking. Indeed, in these theories the scaling of the soft terms is
\eqn\effonescale{\eqalign{& \mu\sim \lambda_u\lambda_d {F\over M}~,\cr & B_\mu\sim \lambda_u\lambda_d {F^2\over M^2}~,\cr & \delta m^2_{{u,d}}\sim \left|\lambda_{u,d}\right|^2{F^2\over M^2}~,\cr & \delta a_{u,d}\sim \left|\lambda_{u,d}\right|^2{F\over M}~.}}
This behaves for all purposes essentially in the same way as \onescale. From the formula for the gaugino masses $m_{\tilde g}\sim \alpha_g {F\over M}$ we conclude that ${F\over M}\gg m_{\tilde g}\sim \mu$. Thus the analogous equation to \constr\ precludes electroweak symmetry breaking.

One imaginable resolution of this specific problem is to engineer field theories in which the scalings \effonescale\ are violated in such a way that $\delta m^2_{{u,d}},B_\mu$ are actually absent at leading order and arise only at higher orders in some small parameter of the hidden-sector.  As we will demonstrate in section 4, such mechanisms are easy to come up with for the $B_\mu$ term as it may carry various charges under PQ and R symmetries, but $m^2_{{u,d}}$ are neutral under all symmetries and it would take a complicated modular structure to achieve this suppression.

For all these reasons we are naturally led to consider genuine two-scale models. These behave differently than the scenarios presented so far.

\subsec{Two (and More) Scale Models}

Here we will study models with two scales such that $F \ll M^2$.  The main new point is that we will assume that $\mu$ is generated in the supersymmetric limit $F=0$.\foot{A conceptually similar assumption regarding electroweak
 symmetry breaking  was discussed in~ \BatraRC.}

According to our counting of powers of $F$, as follows from \wardFonep\ and \dict\ the $B_\mu$ term may arise at linear order in $F$. This does happen in some examples, but contrary to the case of $m^2_{{u,d}}$, since the $B_\mu$ term is sensitive to various global symmetries (and the leading order in $F$ is also subject to holomorphy considerations) it is easy to suppress this contribution. Consequently, we henceforth assume that the leading contribution to the $B_{\mu}$ term comes at most at an order $F^2$. One simple and explicit (symmetry and holomorphy based) mechanism to ensure this will be presented in section 4.
The scalings in this scenario are typically of the form
\eqn\twscale{\eqalign{&\mu\sim \lambda_u\lambda_d M~,\cr & B_\mu\sim \lambda_u\lambda_d {F^2\over M^2}~,\cr &\delta m^2_{{u,d}}\sim \left|\lambda_{u,d}\right|^2{F^2\over M^2}~,\cr  & \delta a_{u,d}\sim \left|\lambda_{u,d}\right|^2{F\over M}~,\cr & m_{\tilde g}\sim  \alpha_g {F\over M}~,}}
where $M$ is the hidden-sector scale. We have also included the typical order of magnitude of gauge mediated contributions in the form of some gaugino mass.

In a large region of parameter space these equations are consistent with viable electroweak symmetry breaking. In general, allowed regions are those of small $\lambda_{u,d}$ and small SUSY breaking.\foot{Small SUSY breaking is easily achieved in many calculable models of SUSY breaking (for several examples, see e.g.\ the review \IntriligatorCP).} This is consistent with the assumptions leading to \twscale.

One particularly interesting case is when the new contributions from the couplings \supgen\ are roughly of the same order of magnitude as the gauge mediated ones. It is easy to see that this happens when
\eqn\same{{F\over M^2}\sim \lambda_{u,d}\sim \alpha_g\sim 10^{-3}-10^{-2}~.}
In this region of parameter space, the contributions from the new couplings do not only give rise to a $\mu$ term but also modify the electroweak breaking parameters significantly. On the other hand, the $\delta a_{u,d}$ terms scale as $\delta a_{u,d}\sim |\lambda_{u,d}|^2{F\over M}$ which means that $a\ll \mu$ and hence they are small and universal in this scenario. Similarly the wrong-Higgs couplings are even smaller than the $A$-terms (because they have further $F/M^2$ suppression).

One could have expected that in order for the contributions from \supgen\ to be comparable to the gauge mediated contributions there would have to be some rough relation between the order of magnitude of  $\lambda$ and the gauge couplings. It is especially encouraging that the couplings to the hidden-sector required in this scenario have garden-variety small values and need not be extremely small or severely tuned. It would have been nicer to derive the relations \same\ from a more complete theory, but we would like to stress that assuming the order of magnitude \same\ is natural both technically and aesthetically.

If the $F$-term in the scenario discussed above is the highest $F$-term in the problem,\foot{That is, we assume that there are no other decoupled fields which have $F$-terms.} the gravitino mass in this scenario comes out to be \eqn\grav{m_{\tilde G}\sim 10{\rm eV}  -  10{\rm keV}~.} Moreover, in any realistic account of low-energy predictions from SUSY breaking, modifications to the formulae of gauge mediation for the Higgs fields should be taken into account but the $A$-terms remain negligibly small in this framework.

To close this subsection we mention that another specific proposal based on two scales involved in generating the Higgs parameters has been discussed in the literature \refs{\RoyNZ,\MurayamaGE} (for more recent works see also~\PerezNG\ and references therein). In essence, the idea is to use a superconformal hidden-sector to suppress the dangerously large terms by appropriately chosen anomalous dimensions. In this scenario the most straightforward estimates of the soft terms are violated due to a peculiar hidden-sector dynamics. It is important to emphasize that to date there is no known superconformal theory which satisfies the necessary inequalities on anomalous dimensions and it will be interesting to see whether such a theory exists.

\newsec{Coupling the Hidden-Sector to the Higgs Fields via an $SU(2)$ Singlet}

In this section we study another possible way of coupling the Higgs fields to the hidden-sector, through an $SU(2)$ singlet superfield $\CS$.\foot{If $\CS$ is an elementary superfield it might develop a large (divergent) tadpole which can destabilize the hierarchy. Symmetries prevent this problem in the examples in section 4.} A familiar example of such a setup is the NMSSM. The superpotential is generally given by
\eqn\cou{
    {\cal W}=\lambda^2{\CS}\CH_u\CH_d+\dots~.}
$\CS$ does not have to be a fundamental singlet, we can think of it as some composite operator in which case the coupling above is non-renormalizable in the UV.

Our zero momentum Lagrangian in components, dropping the terms coming from the K\"ahler potential, is
\eqn\lagcompsing{
    {\cal L}=\lambda^2\left(-\psi_S\psi_{H_u}H_d-\psi_S\psi_{H_d}H_u-S\psi_{H_d}
    \psi_{H_u}+F_SH_uH_d+SF_{H_u}H_d+SH_uF_{H_d}\right)+c.c.~.
}
At leading order we see that the vevs of $S$ and $F_S$ determine
\eqn\muBmusinglet{
    \mu=\lambda^2\langle S\rangle,\qquad B_\mu=\lambda^2\langle F_S\rangle~.
}
Note that at this order in $\lambda$, no other soft terms are generated.  This is different from the case we analyzed in the previous section.

The soft non-holomorphic supersymmetry breaking masses in this theory arise at the order $\lambda^4$ and involve a one-loop computation (we display the 1PI terms of the correlation functions hereafter)
\eqn\softmassloop{\eqalign{    &\delta\langle H_{u,d}H_{u,d}^\dagger\rangle_{p=0}  =-{\lambda^4\over (2\pi)^4}\int d^4q \biggl(\langle\bar\psi_{H_{d,u}}(q)\psi_{H_{d,u}}(-q)\rangle\langle\bar\psi_S(-q)
\psi_S(q)\rangle  \cr & - \langle H_{d,u}(q)H_{d,u}^\dagger(-q)\rangle\langle F_S(-q)F_S^\dagger(q)\rangle-\langle F_{H_{d,u}}(q)F_{H_{d,u}}^\dagger(-q)\rangle\langle S(-q)S^\dagger(q)\rangle\biggr)~.
}}
To leading order in $\lambda$ (and dropping the subleading corrections from the MSSM) we can replace the correlation functions of the Higgs fields by their free values obtaining
\eqn\softmassloop{\eqalign{    &\delta\langle H_{u,d}H_{u,d}^\dagger\rangle_{p=0}\cr &=-{\lambda^4\over (2\pi)^4}\int d^4q \langle H_{d,u}(-q)H_{d,u}^\dagger(q)\rangle \biggl(\sigma^\mu q_\mu\langle\psi_S(q)\bar\psi_S(-q)\rangle -\langle F_S(q)F_S^\dagger(-q)\rangle+q^2\langle S(q)S^\dagger(-q)\rangle\biggr)\cr & =
i{\lambda^4\over (2\pi)^4}\int {d^4q\over 16 q^2} \left\langle \{Q, [Q, \{\bar Q,[\bar Q, S(q) S^\dagger(-q)]\}]\}\right\rangle
~.
}}
Note that $\delta m^2_{u}=\delta m^2_{d}$ in this scenario.
Similarly, the $A$-terms arise from the $HF^\dagger$ two-point functions, which are given by a similar formula
\eqn\softmassloop{\eqalign{    &\langle H_{u,d}F_{H_{u,d}}^\dagger\rangle_{p=0}=-{\lambda^4\over (2\pi)^4}\int d^4q \langle H_{d,u}(q)H_{d,u}^\dagger(-q)\rangle \langle S(-q)F_S^\dagger(q)\rangle\cr & =
-i{\lambda^4\over (2\pi)^4}\int {d^4q\over 4 q^2} \left\langle  \{\bar Q,[\bar Q, S(q) S^\dagger(-q)]\}\right\rangle
~.
}}
Obviously, the two $A$-terms are the same and universal.

To emphasize again, a strikingly different character of a coupling of the form \cou\ compared to the case analyzed in the previous section is that the non-holomorphic masses and $A$-terms arise at different orders in $\lambda$ compared to the $\mu$ and $B_\mu$ term.
One could also discuss the other induced renormalizable couplings, like the wrong-Higgs couplings. It turns out that these arise starting from two-loop diagrams. (They are given by double integrals over certain combinations of three-point functions of the hidden-sector.)

\subsec{One-Scale}

If the hidden-sector is a single-scale theory, the vevs of $S$ and $F_S$ as well as all the correlation functions are fixed by dimensional analysis:
\eqn\scal{\eqalign{& \mu\sim \lambda^2M~,\cr
& B_\mu\sim \lambda^2 M^2~,\cr
& \delta m^2_{{u,d}}\sim {\lambda^4\over 16\pi^2}M^2~,\cr
& \delta a_{u,d}\sim {\lambda^4\over 16\pi^2}M~.}}

Thus, this scenario predicts a ratio which is too large between $B_\mu$ and $\mu$. One may imagine some ways to tame the large $B_\mu$ term, in which case this scenario becomes viable. Note that unlike the one-scale scenario in the case of coupling the Higgs fields to hidden-sector doublets, $\delta m^2_{{u,d}}$ are small in this framework and are not problematic. These are the ones which are typically harder to suppress (as they are neutral under all symmetries), while it is rather easy to come up with mechanisms to suppress the $B_\mu$ term as we shall see in section 4.

The case in which the $\mu$ term is generated due to SUSY breaking is effectively obeying the same relations as a one-scale model although it really has two scales $\sqrt F,M$. Indeed, in this case we typically get
\eqn\scalfm{\eqalign{& \mu\sim \lambda^2{F\over M}~,\cr
& B_\mu\sim \lambda^2 F~,\cr
& \delta m^2_{{u,d}}\sim {\lambda^4\over 16\pi^2}{F^2\over M^2}~,\cr
& \delta a_{u,d}\sim {\lambda^4\over 16\pi^2}{F\over M}~.}}

Obviously, here we encounter again that the ratio of $B_\mu$ to $\mu$ is too large. Note that a further suppressed $B_\mu\sim \lambda^2 {F^2\over M^2}$ is not enough.  We conclude that to realize this scenario in a phenomenologically viable way, the $B_\mu$ term has to be suppressed presumably by additional powers of some small hidden-sector Yukawa couplings.  This framework is precisely the one in which the the $\mu/B_\mu$-problem is often discussed and some explicit solutions have been proposed\ \refs{\DvaliCU,\DineXK
\GiudiceCA\YanagidaYF\DimopoulosJE\LangackerHS\HallUP-\LiuPA}. These solutions rely on the following fact.  Since the $\mu$ term in \scalfm\ generally arises at the one-loop order of the hidden-sector, this means that one has to ensure that the $B_\mu$ term is generated at two loops or higher order.  This is not easily achieved, so one has to engineer hidden-sector theories with several components which communicate weakly.

\subsec{Two Scales}

Having seen that one-scale models are hard to build we turn to models with two scales, one associated to a mass scale $M$ and the other is the scale of SUSY breaking $\sqrt F$. In analogy with our discussion in the previous section, we propose to consider models in which the $\mu$ term is generated in the SUSY limit while the $B_\mu$ term, as well as all the other soft terms, are generated by SUSY breaking. As in the previous section, there exist simple ways to suppress the leading large contribution to the $B_\mu$ term. For instance, unlike in \scalfm, if $B_\mu$ arises at order $F^2/M^2$, we can find realistic electroweak symmetry breaking. Furthermore, as we will see in the next section, it is easy to come up with other mechanisms to suppress the $B_\mu$ term without introducing complicated multi-component hidden-sectors.

Having assumed that, we are led to the naive estimates
\eqn\twoscal{\eqalign{& \mu\sim \lambda^2M~,\cr
& B_\mu\ll \lambda^2 F~,\cr
& \delta m^2_{{u,d}}\sim {\lambda^4\over 16\pi^2}{F^2\over M^2}~,\cr
& \delta a_{u,d}\sim {\lambda^4\over 16\pi^2}{F\over M}~.}}

As in the previous section, we see that that these models lead to electroweak symmetry breaking if the couplings $\lambda_{u,d}$ are small (validating our analysis). In some realizations one also has to require that SUSY breaking is small $F\ll M^2$, but not always. In section 4 we will see an example where $B_\mu/\mu\sim m_{soft}$ is guaranteed by the dynamics of the model and one could typically have $\lambda^2\sim 10^{-2}$.

The models \twoscal\ are interesting because they give rise to a universal prediction. Since $\lambda$ is a small parameter, $\delta m^2_{{u,d}}$  are always negligibly small which means that the gauge mediated contributions to $m^2_{u,d}$ dominate. So, we expect the Higgs soft masses to be virtually identical to the slepton-doublet mass (at the boundary scale of the RG flow)
\eqn\lepdou{m_{{u}}^2=m_{{d}}^2=m_{\tilde L}^2~. }

Consequently, the two different ways to generate $\mu$ and $B_\mu$ have rather different possible signatures. The one summarized in \twscale\ provides a space of possibilities with $\delta m^2_{{u,d}}$ potentially comparable to the gauge mediated contributions.  The mechanism based on a singlet coupling to $\CH_u\CH_d$ \cou\ predicts that all the non-holomorphic mass terms are dominated by gauge mediation. On the other hand, it is a common feature of both situations that the $A$-terms and the hard terms are negligibly small.

\newsec{Explicit Models Based on Messengers}

So far our discussion has been very general.  It applies to many theories including models of dynamical supersymmetry breaking where all the dimensionful scales are dynamically generated.  In this section we will consider explicit examples. These will be weakly coupled theories with canonical K\"ahler potential and explicit dimensionful parameters.  We can view them as effective theories of other, more complete theories.  The purpose of studying these examples, apart from their own interest, is to provide existence proofs of our general discussion.

We will investigate models of messengers which have R-symmetry and can generate the Higgs parameters. Models of messengers with R-symmetry were studied in the context of gauge mediation in~\CheungES\ and were shown to possess certain appealing features. Here we will see that R-symmetry is useful for the Higgs-sector model building too.

\subsec{Hidden-Sector Doublets}

Consider a general model of messenger fields arranged in column vectors $\eta$ and $\tilde\eta$. Let us suppose that these messengers are coupled to the Higgs fields as follows
\eqn\messengerscou{
    {\cal W}=M_0\left(\lambda_u\CH_u\eta^Tc+\lambda_d\CH_d\tilde c^T\tilde\eta\right)+\eta^TM\tilde\eta+\CX\eta^T\lambda\tilde\eta~,
}
where $c,\tilde c$ are some column vectors (which project on $SU(2)$ doublets) and $M,\lambda$ are matrices. $M_0$ is a mass scale roughly of the order of entries in the matrix $M$. To respect messenger parity we impose that the matrices $M,\lambda$ are symmetric (up to a unitary transformation).\foot{We thank Michael Dine for a helpful comment about messenger parity.}

In general one can imagine that such a theory arises in the low energy description of some dynamical setup where the operators $\Phi_u$ and $\Phi_d$ are dimension two operators in the UV undergoing dimensional transmutation and become $M_0\eta^Tc$ and $M_0\tilde c^T\tilde\eta$, where $\eta$ and $\tilde \eta$ are dimension one fields at low energies. In this context, the absence of a tree-level $\mu$ term is either guaranteed using some high energy symmetries or can be simply assumed to be generated only due to non-perturbative dynamics.

We will denote \eqn\defin{{\cal M}=M+\CX\lambda~.}
$\CX$ is imagined to be some hidden-sector superfield whose bottom component and $F$-component have vevs
\eqn\X{
    \CX= X +\theta^2 F~.
}
Classically integrating-out all the messenger fields is straightforward.
The resulting superpotential and K\"ahler potential are
\eqn\supeffe{
    {\cal W}_{eff}=-M_0^2\lambda_u\lambda_d\tilde c^T{\cal M}^{-1}c\CH_u\CH_d~,
}
\eqn\kahleffe{
    {1\over |M_0|^2}\delta{\cal K}_{eff}=|\lambda_d|^2\CH_d\CH_d^\dagger\Tr\left(({\cal M}^{-1})^*({\cal M}^{-1})^T\tilde c\tilde c ^\dagger\right)+|\lambda_u|^2\CH_u\CH_u^\dagger\Tr\left(({\cal M}^{-1})^\dagger {\cal M}^{-1} c c ^\dagger\right)~.
}

Now we can find the $\mu$ and $B_\mu$ terms and the other soft terms. Let us begin with the contributions arising from the superpotential.  The result is directly read from \supeffe\ and it is given for small $F$ by
\eqn\mumesse{
    -\mu=M_0^2\lambda_u\lambda_d\Tr\left({1\over M+X\lambda}c\tilde c^T\right)~,
}
\eqn\Bmu{
    B_\mu=F{\partial\over \partial X}\mu(X)=M_0^2\lambda_u\lambda_d F \Tr\left({1\over M+X\lambda}\lambda{1\over M+X\lambda}c\tilde c^T\right)~.
}
Note that $\mu$ is generated even for $F=0$.  Since the right hand side of \Bmu\ is non-zero in generic models of messengers, we get $B_\mu\sim F$. This is too large.

The leading order contribution to $B_\mu$ in \Bmu\ can be easily excluded if we assume that the couplings \messengerscou\ respect an R-symmetry such that $R(H_u)+R(H_d)=2$ and $R(\CX)\not=0$.  These conditions (and holomorphy) guarantee that $\mu(X)$ of \mumesse\
is independent of $X$.  Therefore, $B_\mu$ in \Bmu\ vanishes at the leading order.  Note, that here we used a selection rule due to a $U(1)_R$ symmetry even though it is spontaneously broken due to the nonzero vev $X$.

The corrections to the K\"ahler potential in \kahleffe\ give rise to the soft non-holomorphic masses and $A$-terms.  Other corrections due to higher derivatives or quantum effects lead to a negligible $B_\mu$.

Additional curious feature of \messengerscou\ (but not of the entire class) is that $\delta m^2_{{u,d}}$ turn out negative. This can be seen directly from \kahleffe, or by using the more general argument in \EGGM.  From the phenomenological point of view this means that in these models the soft non-holomorphic Higgs masses are always smaller than the gauge mediation value.

In order to infer the gaugino masses we have to study the determinant of the mass matrix, $\det({\cal M})$. The conclusion, as explained in~\CheungES, is that if
\eqn\gaugino{\sum_{i}\left(2-R(\eta_i)-R(\tilde \eta_i)\right)\neq 0 ~,}
the gaugino masses are expected not to vanish at leading order in $F$.

To summarize, R-symmetric models of messengers which couple quadratically to the Higgs fields and satisfy the simple conditions
\eqn\conditiosn{R(\CX)\not=0~,\qquad \sum_{i}\left(2-R(\eta_i)-R(\tilde \eta_i)\right)\not=0~,\qquad R(H_u)+R(H_d)=2}
lead to viable models of the type discussed around \twscale, sharing all the general features of this class of models, but also give rise to more special traits. (Especially, a negligible $B_\mu$ term and negative $\delta m^2_{{u,d}}$.)
The typical scalings of soft terms in these models are
\eqn\scamode{\mu\sim \lambda_u\lambda_d M,\qquad B_\mu=0,\qquad \delta m^2_{{u,d}}\sim |\lambda_{u,d}|^2{F^2\over M^2},\qquad \delta a_{u,d}\sim |\lambda_{u,d}|^2{F\over M}~, }
and we also know that
\eqn\sign{\delta m^2_{{u,d}}\leqslant 0~.}

\subsec{Hidden-Sector Singlet}

Consider a theory of the form discussed in \cou\
\eqn\higgssup{
    {\cal W}=\lambda^2 \CS\CH_u\CH_d+{\cal W}_{hidden}~.
}
In this subsection we will discuss two simple hidden-sector theories described by the superpotentials \foot{These models closely resemble those of (E)OGM \CheungES. The precise relation is that we have promoted some of the mass parameters of (E)OGM to a chiral superfield $\CS$. }
\eqn\hiddensectorone{{\cal W}_{hidden}^{(1)}= \CX(\sum_{i=1}^3\rho_i\tilde\rho_i+\eta\tilde\eta)+ \CS\rho_1\tilde\rho_2+M\rho_2\tilde\rho_3~,}
\eqn\hiddensectortwo{{\cal W}_{hidden}^{(2)}= \CX(\sum_{i=1}^4\eta_i\tilde\eta_i)+ \CS\left(\eta_1\tilde\eta_2+\eta_3\tilde\eta_4\right)+M\eta_2\tilde\eta_3~,}
where  $M$ is some mass scale and as above
 \eqn\vevandf{\CX= X+\theta^2 F}
is fixed.\foot{One could have also studied the effective potential for $X$ and search for models in which it is stabilized. (This comment is relevant for the previous subsection as well.) In this way we would remove the need for assuming the vev and $F$ term of $\CX$, but for simplicity we do not perform an exhaustive analysis of this possibility here.}
The superfields $\rho$ and $\tilde\rho$ in \hiddensectorone\ are neutral under the gauge symmetries, while the messengers $\eta$ and $\tilde\eta$ in \hiddensectorone,\hiddensectortwo\ transform as ${\bf 5 \oplus \bar 5}$ of $SU(5)$. It is important to remark that we concentrate on the specific models \hiddensectorone,\hiddensectortwo\ for simplicity.  Our conclusions and analysis apply to more general models of messengers. Note that both theories are generic for some $U(1)_R$ symmetry and a set of $U(1)$'s.  The existence of an R-symmetry is important for the mechanism described below to operate. It is also important to note that both theories \hiddensectorone\ and \hiddensectortwo\ respect messenger parity.  Finally, these models clearly generate all the usual soft masses via gauge mediation.

Our purpose is to study the dynamics of the $\CS$ superfield. $S$ is a pseudo-modulus and $F_S=0$. The idea is that these tree-level properties allow for $S$ to acquire large vev (of order $M$) once quantum corrections are incorporated. This vev is not suppressed by the loop expansion parameters. However, the expectation value of $F_S$ is proportional to the loop expansion parameter and therefore it is naturally much smaller than the largest $F$-term in the problem \vevandf.

Once the messengers are integrated out, an effective K\"ahler potential is generated~\GrisaruVE\
\eqn\effeckah{
    {\cal K}_{eff}=\CS\CS^\dagger-{1\over 32\pi^2}\Tr\left[{\cal M}{\cal M}^\dagger\ln\left({\cal M}{\cal M}^\dagger/\Lambda^2\right)\right]~.
}
We have denoted by ${\cal M}$ the total mass matrix appearing in the superpotential.

In components this Lagrangian takes the form
\eqn\auxLag{
    {\cal L}=\left(1+\partial_{X,X^\dagger}\delta{\cal K}_{eff}\right)FF^\dagger+\left(1+\partial_{S,S^\dagger}\delta{\cal K}_{eff}\right)F_SF_S^\dagger+\left(\partial_{X,S^\dagger}\delta{\cal K}_{eff}F
    F_S^\dagger+c.c.\right)~.
}
To leading order in the loop expansion the equation of motion of $F_S$ is
\eqn\solution{
F_S=-F\partial_{X,S^\dagger}\delta{\cal K}_{eff} ~,
}
which leads to a one-loop generated $F$-term for $\CS$.

The resulting Higgs parameters from this general mechanism are
\eqn\scalings{
     \mu=\lambda^2\langle S\rangle \sim \lambda^2 M,\qquad B_\mu=\lambda^2\langle F_S\rangle \sim \lambda^2 {F\over 16\pi^2}~.
   }
Note that ${B_\mu \over \mu}\sim {1\over 16\pi^2}{F\over M}$ comes out {\it automatically} at the electroweak scale. This model does not require decoupled sectors with complicated interactions to solve the $\mu/B_\mu$-problem, rather, once we assume the $\mu$ term arises due to supersymmetric dynamics the model building simplifies dramatically. Lastly, as explained in section 3, $m^2_{u,d}$ are dominated by the gauge mediation contribution.

For the sake of having a numerical example, consider $\langle X\rangle= M$. Then, the minimum of the potential of the first model \hiddensectorone\ is at $S\approx 0.40 M$ with $F_S\approx {0.8\over 32\pi^2}F$, and the minimum of the potential of the second model \hiddensectortwo\ is at $S\approx 0.43 M$ with $F_S\approx {9\over 32\pi^2}F$.

\bigskip

\noindent {\bf Acknowledgments:}

We would like to thank N.~Arkani-Hamed, M.~Buican, M.~Dine, A.~Katz, P.~Meade, Y.~Nir, D.~Shih, S.~Thomas, and T.~Volansky for useful discussions. The work of ZK was supported in part by NSF grant PHY-0503584 and that of NS was supported in part by DOE grant DE-FG02-90ER40542. Any opinions, findings, and conclusions or recommendations expressed in this material are those of the author(s) and do not necessarily reflect the views of the funding agencies.

\listrefs
\end